\documentclass[12pt]{article}
\input epsf.sty
\def\be{\begin{equation}} \def\ee{\end{equation}}
\def\bea{\begin{eqnarray}} \def\eea{\end{eqnarray}} \def\ba{\begin{array}}
\def\ea{\end{array}} \def\ben{\begin{enumerate}} \def\een{\end{enumerate}}

\newcommand{\eqn}[1]{(\ref{#1})}

\newcommand{\hepth}[1]{{\tt hep-th/{#1}}}

\def\pa{\partial}

\def\br{\nonumber\\}

\begin{document}
{}~
\hfill\vbox{\hbox{hep-th/yymm.nnnn} \hbox{\today}}\break

\vskip .5cm
\centerline{\large \bf
 M2-branes on a resolved $C_4/Z_4$ }
\vskip .5cm

\vspace*{.5cm}

\centerline{  Harvendra Singh}

\vspace*{.25cm}
\centerline{ \it Theory group, Saha Institute of Nuclear Physics} 
\centerline{ \it  1/AF, Bidhannagar, Kolkata 700064, India}
\vspace*{.25cm}

\vspace*{.5cm}

\vskip.5cm
\centerline{E-mail: h.singh[AT]saha.ac.in }

\vskip1cm
\centerline{\bf Abstract} \bigskip

We write down M2-branes  on the resolved $C_4/Z_4$ orbifold space.
The resolved spatial geometry is such that it interpolates between 
$R^2\times CP_3$ near the branes
and $C_4/Z_4$ asymptotically. 
The near horizon geometry of these branes is a deformation of $AdS_4\times 
S^7/Z_4$. An interesting aspect is that for  $k=4$  
Chern-Simons theory  coupling becomes vanishing 
near the IR cutoff leading to spontaneous compactification to type IIA.

\vfill \eject

\baselineskip=16.2pt

\section{Introduction}

The study of spacetime solutions which represent extended 
$p$-brane  objects has led to many new ideas in string theory as 
well as in field theory \cite{maldacena,aharony}. Interestingly, very
recently 
there has been a  fast activity in this field, see [3-42], where
certain type of Chern-Simons field theories in 
$2+1$ dimensions have been 
proposed to be dual to M-theory on $AdS_4\times S^7/Z_k$ 
spacetime. Namely, the Aharony-Bergman-Jafferis-Maldacena (ABJM) 
Chern-Simons theory \cite{abjm}, which has 
${\cal N}=6$ 
$SU(N)\times SU(N)$ superconformal 
symmetry, is conjectured to be dual to  M-theory on $AdS_4\times S^7/Z_k$ 
with level $k>2$. While the 
originally proposed Bagger-Lambert (BL) membrane theory based 
on compact tri-algebras has maximal ${\cal N}=8$ 
superconformal symmetry but is known only for $SO(4)\times SO(4)$ 
R-symmetry \cite{bl,Gustav}. 
Although by allowing noncompact tri-algebras, BL theory can be extended to 
admit full $SU(N) 
\times SU(N)$ symmetry \cite{verlinde}. But these theories have a 
ghost field in the spectrum which when gauge-fixed to a constant 
value gives rise to $SU(N)$ superconformal Yang-mills theory
\cite{Bandres:2008kj}. These 
developments are necessary
to understand the M-theory origin of superconformal $SU(N)$ Yang-Mills 
gauge theory which lives on the D2-branes over $AdS_4\times S_6$, and 
vice-versa.     

For the current 
purpose the paper is organised as follows. 
In the section-2, we 
review basic properties of the resolved $C_4/Z_k$ orbifold geometry
and write down the M2-brane solution on the resolved space. In 
section-3 we discuss spontaneous compactification to type IIA background.
We discuss the nature of the singularity at the IR cut off scale. 
Near the IR cutoff the string (Chern-Simons) coupling vanishes but 
curvature also become large. 
The results are summarised in the 
last section.  

 \section{M2 on resolved $C_4/Z_k$ }
The  flat metric on $C_4/Z_k$  eight-dimensional space can be 
written as
 \be\label{sch}
ds_{C_4/Z_k}^2={dr^2}+ {r^2\over k^2} (dz+k 
A)^2  +{r^2} ds^2_{CP_3}
\ee
where $r^2= (y^m)^2 $. The $y^m$'s are eight 
Cartesian coordinates
which define the base space of the geometry, that is 
$R^{8}$ or $C_4$. 
The round Fubini-Study metric on unit size $CP_3$ space can be read from 
\cite{cveticcp3,
abjm} and it is
\bea
ds_{CP_3}^2=d\xi^2 + \cos^2\xi\sin^2\xi~ (\tilde\psi)^2
+ {\cos^2\xi\over4}(d\theta_1^2+\sin^2\theta_1 d\phi_1^2) +
{\sin^2\xi\over4}(d\theta_2^2+\sin^2\theta_2 d\phi_2^2) \br
\eea
where the coordinate ranges are $0\le\xi<{\pi\over2},~0\le 
z<2\pi,~0\le\theta_i<\pi,~0\le\phi_i<2\pi$. The $\tilde\psi$ and the 
1-form along the Hopf fibre $z$ are given as
\bea 
&& \tilde\psi\equiv d\psi 
+{\cos\theta_1\over2}d\phi_1
-{\cos\theta_2\over2}d\phi_2 
\br && A={1\over2}\left((\cos^2\xi-\sin^2\xi)d\psi+\cos^2\xi 
\cos\theta_1d\phi_1
+\sin^2\xi\cos\theta_2d\phi_2\right)
 \eea
The space is asymptotically locally Euclidean (ALE)  but has (orbifold) 
conical singularity at 
$r=0$ for all $k\ge2$.
 
The  M2-brane solution on this transverse space is given by
\bea\label{m2}
&&ds_{11}^2=h^{-{2\over3}} (-dx_0^2+dx_1^2+dx_2^2) + h^{1\over3} 
ds^2_{C_4/Z_k}
\br &&F_{(4)}=d(h^{-1})\wedge dx^0\wedge dx^1\wedge dx^2
\eea
where the harmonic function is 
\be
h(r)=1 +{2^5 \pi^2 Nk l_p^6\over r^6}. \label{m2f}\ee
The background preserves $3/8$ supersymmetries.
The near horizon limit $(r\to l_p^2 U,~l_p\to 0)$ of \eqn{m2}
gives us M2-branes 
on $AdS_4 \times S_7/Z_k$ spacetime
\bea\label{m2b}
&&ds_{11}^2\sim{R^2} (U^4(-dx_0^2+dx_1^2+dx_2^2)+{dU^2\over U^2}) + 
R^2 ds^2_{S_7/Z_k}
\br &&F_{(4)}\sim{6R^3}~ vol(AdS_4)
\eea
where $(R/l_p)^2=(2^5\pi^2 Nk)^{1/3} $. Here $l_p$ is the 
eleven-dimensional Planck length.

The number of M2-branes in \eqn{m2} is taken as $N\cdot k$, so that 
the flux through 
$S^7/Z_k$ remains integral of $N$. 
The doubling of supersymmetries in the near horizon region suggests that 
the $ AdS_4$ geometry will
preserves 24 supersymmetries. The holographic dual boundary Chern-Simons 
theory in large $N$ 
($k>2$) limit is conjectured to be the ${\cal N}=6$ $SU(N)_k\times 
SU(N)_{-k}$ superconformal Chern-Simons field theory living on the 
worldvolume of $Nk$
M2-branes \cite{abjm}. While in the large $N$ 't 
Hooft limit, but 
with fixed $N/k$ ratio,
 the  theory reduces to weakly coupled superconformal
Chern-Simons theory of corresponding $N$ D2-branes 
on $AdS_4\times CP_3$ \cite{abjm}.   

\subsection{Special case of $C_4/Z_4$}
Our interest is in the special case of orbifold space $C_4/Z_4$ where we 
are able to resolve the conical 
singularity at the origin. The modified metric on $C_4/Z_k$ is 
taken as  Eguchi-Hanson type \cite{eh},
 \be\label{sch1}
ds_{C_4/Z_4}^2={dr^2\over f(r)}  + {r^2\over k^2} 
{f(r)}(dz+ k A)^2 +{r^2} ds^2_{CP_3}
\ee
which can be solved exactly for Ricci flatness. We determine that there is 
a unique solution  for any $k$ value 
\be
f(r)= (1-r_0^8/r^8)
\ee
where $r_0$ is an integration constant. The  coordinate ranges 
are  fixed as \be\label{ranges1}
r_0\le r\le\infty,~~0\le z \le 2\pi .
\ee
To know if the metric is regular
near $r=r_0$ region, we can define a local coordinate patch 
$$r^2 (1-r_0^{8}/r^{8})=(k\rho)^2$$ 
with $\rho$ being infinitesimal radial coordinate. The $r=r_0$ 
neighborhood geometry then becomes
\be\label{kl2}
ds^2\simeq {k^2\over 16}d\rho^2 + \rho^2 (dz+kA)^2 +r_0^2 
ds^2_{CP_3}
\ee
So the metric will be resolved only when $k=4$. 
For $k=4$ only, eq.\eqn{kl2}  will have a smooth $R^2 \times CP_3$ 
geometry in 
the neighborhood of $r=r_0$.
This makes the fibered 
circle $z$ to be well behaved everywhere. One can think of the geometry 
in \eqn{kl2} as if every point in the 
$CP_3$ has a small $R^2$ patch attached, while the $CP_3$  space has a  
large 
but  constant radius given by $r_0$.\footnote{The resolved manifold is 
topologically a complex line bundle over $CP_3$ \cite{cvetic}. } This 
resolution is 
similar in manner 
as to the resolved  Eguchi-Hanson $4D$ instantons \cite{eh} and the 
resolution of 
Calabi-Yau cones in higher dimensions, see \cite{zt,ks,cvetic1,hs,km}. For 
a 
detailed study of  deformations and resolutions of various Calabi-Yau 
spaces 
in higher dimensions with fluxes turned on one can see \cite{cvetic}. 
\footnote{We comment that one can anyway study 
M2-branes on the transverse space in \eqn{sch1} for any $k$ so long as we 
do not bother about the orbifold singularity at $r=0$. Actually the 
space patch inside $r=r_0$ no longer exists, since $r\ge r_0$. But in the 
$\rho$ coordinate which is appropriate coordinate in the $r=r_0$ region 
the singularity will exist which can be seen from 
\eqn{kl2}.} 

\noindent{\it M2-brane solution}

Correspondingly the $4N$ branes background on resolved $C_4/Z_4$ space can 
be obtained  by solving
$$ \partial_r r^7 f \partial_r h=0 $$
The complete solution is as in equation \eqn{m2} but with a new harmonic 
function 
\be
h(r)= 1
 + {Q\over 4r_0^6}  \left(\arctan({r^2\over r_0^2})
 - {1\over 2} \log{(r^2 - r_0^2)\over (r^2 + r_0^2)}\right) 
\ee
Near $r=r_0$ tip, this harmonic function behaves as 
$$ h\simeq {Q\over 8 r_0^6}\log({r_0^2\over 2\rho^2}) .$$
So the solution  diverges but logarithmically only. 
While for $r\gg r_0$ it 
becomes $$ h\sim 1+ {Q\over 6 r^6} + {Qr_0^8\over 14r^{14}}+ \cdots$$
Comparing it with \eqn{m2f} we  fix $ {Q/6}\equiv 2^5 \pi^2 4N l_p^6$.

The near horizon decoupled geometry in this case is
\bea\label{m2a}
&&ds_{11}^2\simeq l_p^2 h^{{1\over3}}U^2\left( 
{(-dx_0^2+dx_1^2+dx_2^2)\over 
U^2h} + 
{dU^2\over U^2 f} +{f\over 16}(dz+4A)^2+
ds^2_{CP_3} \right)
\br &&F_{(4)}=l_p^{3}d(h^{-1})\wedge dx^0\wedge dx^1\wedge dx^2
\eea
with 
\be\label{nk1}
h(U)\simeq 
 {Q_0\over 4U_0^6}\left({1\over 2} \log{(U^2 + U_0^2)\over (U^2 - U_0^2)} 
 +   \arctan({U^2\over U_0^2})\right)
\ee 
and $f=1-{U_0^8\over U^8}$. Here we identify $ {Q_0\over 6}\equiv 2^5 
\pi^2 (4N) \gg 
1$ so that the overall curvature of the spacetime is small in the Planck 
units.
It is obviously a difformation of the $AdS_4\times S^7/Z_4$ discussed 
above as it can be seen that in the far UV regime ($U\gg U_0$) the 
geometry in \eqn{m2a} becomes exactly  
\bea\label{m2b1}
&&ds_{11}^2\sim l_p^2 (Q_0/6)^{{1\over3}}\left( 
{U^4(-dx_0^2+dx_1^2+dx_2^2)\over 
(Q_0/6)} + 
{dU^2\over U^2 } +{1\over 16}(dz+4A)^2+
ds^2_{CP_3} \right)
\br &&F_{(4)}\sim l_p^3{36\over Q_0}U^5 dU\wedge dx^0\wedge dx^1\wedge 
dx^2
\equiv l_p^3\sqrt{6Q_0} ~vol(AdS_4)
\eea
which is the near horizon geometry $AdS_4\times 
S^7/Z_4$ corresponding to  $4N$ M2-branes on unresolved $C_4/Z_4$.

\section{D2-branes}
In order to study corresponding type IIA string picture we need to 
compactify along the fibre $z$ in $S^7/Z_4$. As the  
 $U$ decreases, at some value the effective radius of the 
$z$ circle 
will become smaller than the eleven-dimensional Planck length and we have 
to think in terms 
of type IIA strings. We can then 
compactify along $z$ and the corresponding ten-dimensional metric and 
dilaton are obtained from \eqn{m2a}\footnote{ Our convention is 
$(l_p /R_{(11)})^3=1/g_s^2$ . }
\bea \label{st2}
&& ds_{str}^2= e^{2\phi\over3}h^{1\over3} U^2
\left( {(-dx_0^2+dx_1^2+dx_2^2)\over 
hU^2} + {dU^2\over U^2 f} +ds^2_{CP_3} \right)\br
&& e^{4\phi\over3}= h^{{1\over3}}{fU^2 \over 16} 
\eea
where $h$ is as in \eqn{nk1} and $f=1-U_0^8/U^8$ and string length is set 
to one. This should be thought 
of as a background 
due to $N$ D2-branes on deformed $AdS_4\times CP_3$ in 
the IR region where string coupling is weak. Only in the far UV limit $(U\gg 
U_0)$ or when $U_0=0$ 
 we will get $AdS_4\times CP_3$. Specially, near $U=U_0$ 
(cut-off) IR region strings 
become essentially non-interacting as $f$ vanishes there, however the 
radius of curvature given by  $e^{\phi\over3}h^{1\over6} U$ 
also
becomes small at the same time. So there is curvature singularity 
which will require higher order $\alpha'$ corrections to the 
string geometry.
In order to get actual IR behavior we define $(1-U_0^8/U^8)=(4u)^2$
in the neighborhood of $U=U_0$ which is a chosen lowest energy scale in 
our theory. In this neighborhood the radius of curvature, $R_{(10)}$, and 
the string 
coupling 
behave as
\bea\label{coupl}
&& (R_{(10)})^2\sim x \sqrt{\log{1\over 2x^2}} \sqrt{4N} \br
&& e^{2\phi}\sim x^3 \sqrt{\log{1\over 2x^2}} \sqrt{4N}
\eea
where $x\equiv {u\over U_0}$. 
As $x\to0$ it can be seen that string coupling vanishes  faster than the 
curvature radius of the string metric. It means 
that the strings become non-interacting. However higher order world-sheet 
corrections have to be included in order to know the dynamics at the IR 
cutoff.   We have plotted this behavior of quantities in eqs.\eqn{coupl} 
in the graph 
below.
\begin{figure}[!ht]
\leavevmode
\begin{center}
\epsfysize=6cm
\epsfbox{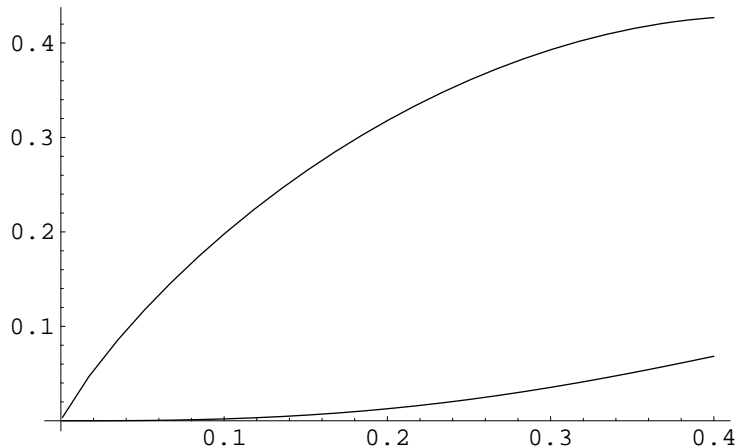}
\end{center}
\caption{\it 
This represents the plot of $(R_{(10)}/l_s)^2$ (upper) and $e^{2\phi}$ 
(lower) Vs 
$x$. The string coupling vanishes faster. 
} \label{fig.6}
\end{figure}

\section{World-volume theory}

We shall now comment on the  $N=6$ world-volume theory of the 
M2-branes on the resolved $AdS_4\times S^7/Z_4$ for finite level $k=4$. 
As we have noted above the $N=6$ Chern-Simons theory always 
flows to the weak coupling in $IR$ for $k=4$ being finite. 
We recover the D2-brane theory on $AdS_4\times CP_3$ in the IR. 

 In the 
BL theory there is an additional scalar field which is considered to be a 
free field. Its 
vev controls the 
strength of the $3D$ gauge coupling. In the paper \cite{shamik}, it 
was 
shown that there is such a field which represents the center of mass 
scalar field of M2-branes corresponding to the 
location of the branes in the flat transverse space. 
We would like to see whether this survives the interpretation once 
the transverse $C_4/Z_4$ space is resolved or is no longer flat 
Euclidean.

Here we just write down the Born-Infeld action.
The world-volume metric on M2-brane in flat Euclidean transverse space 
can be written as  
\bea
&& G_{\mu\nu}\equiv \eta_{\mu\nu}+ \sum_{M=1}^8 G_{MN}(Y)\pa_\mu 
Y^M\pa_\nu Y^N
\eea
where $(\mu,\nu=0,1,2)$ and $(M,N=3,4,\cdots,10)$.
So it has an explicit $SO(8)$ invariance.
Now identifying the $Y^{10}\equiv z$ and defining $G_{10\,10}=e^{4\phi/3}$ 
and 
taking all $G_{MN}$ independent of $z$ field, we can write  
\be
-\int d^3 x \sqrt{Det(G)}= -\int d^3 x e^{-\phi} \sqrt{|g_{\mu\nu} 
+g_{mn}
\partial_\mu Y^m \partial_\nu Y^n+ e^{2\phi} 
\partial_\mu z \partial_\nu z|}  
\ee
where the scalars $Y^m,~z$ constitute the all 8 scalars and $g$ is 
the string metric. Doing this spontaneously breaks the overall $SO(8)$ 
symmetry to 
$SO(7)\times U(1)$. 
We can also see that in the limit when string coupling vanishes, the 
$z$ kinetic
terms become subleading and could be dropped, leaving behind DBI action 
for D2-branes. Writing down the full action for the 
background  with a resolved transverse space metric 
 \be\label{sch2}
ds_{C_4/Z_4}^2={dr^2\over f(r)}  + {r^2\over k^2} 
{f(r)}(dz+ k A)^2 +{r^2} ds^2_{CP_3}
\ee
we get an effective BI action (for $K=4$)
\be
 -\int d^3 x  \sqrt{|\eta_{\mu\nu} 
+{1\over f(R)}
\partial_\mu R \partial_\nu R+  {R^2\over 16} f(R) 
(\partial_\mu z + 4A_\mu)(\partial_\nu z +4A_\nu)+ \cdots|}  
\ee
where $f(R)=1-r_0^8/R^8$. Thus we can see that  
there 
can be an effective  potential for 
scalar field $R$ once $z$ is Higgsed by gauge field $A$, however it can be 
a free modulus only if $dz+4A$ is vanishing. While $z$ as it appears 
only through derivatives will be a free field. The gauge fields in the 
above 
do not have kinetic terms (as those are not dynamical) but they can have 
Chern-Simons like terms. 
This indicates that the complete Chern-Simons theory would be a
gauged version of $N=6$ ABJM 
theory (for $k=4$) with appropriate superpotential for scalar fields.
It will be interesting to know such a theory for $k=4$.

\section{Summary}
In this short note we have constructed M2-brane solutions  on 
resolved $C_4/Z_4$ 
Euclidean 8-manifolds. These solutions possess new 
properties which to our knowledge have not been explored earlier in 
the literature. The near horizon geometry is a deformation of the 
$AdS_4\times S^7/Z_4$ spacetime of the 
M2-branes. Although there is essential singularity in the IR region 
where curvatures become large but the string (Yang-Mills) coupling 
vanishes. This is juxtaposite of the superconformal Yang-Mills theories 
for which have strong coupling fixed point in the IR. 
We conclude that the
 holographic dual $SU(N)\times SU(N)$ Chern-Simons theory is 
free near the IR cutoff. Although in IR the curvature of spacetime becomes 
very small but the string coupling vanishes there. In other words
 the core of the resolved M2-branes
dissolves into noninteracting D2-branes and the string 
coupling vanishes.  Since the IR 
region has an inbuilt cutoff $U=U_0$ near which curvature becomes high and 
therefore higher order worldsheet corrections should be taken into 
account.  
We have also commented that the complete Chern-Simons theory would be a
gauged version of $N=6$ ABJM 
theory (for $k=4$) with appropriate superpotential for scalar fields.

\section*{Acknowledgements}
I am grateful to Sunil Mukhi for his useful comments on the draft. I 
also wish 
to acknowledge the warm hospitality during the "Monsoon Workshop on 
String Theory" at ICTS-TIFR, Mumbai. The author is also  an
Associate  of
AS-ICTP, Trieste where this work was mainly carried out and this support 
is highly acknowledged.


\begin{thebibliography}{99}

\bibitem{maldacena} J.M. Maldacena, {\it The large N limit of 
superconformal field theories and supergravity}, Adv. Theor. Math. Phys. 2 
(1998) 231, \hepth{9711200}; S.S. Gubser, I.R. Klebanov, A.M. Polyakov, 
Phys. Lett. 
428B (1998) 105, \hepth{9802109}; E. Witten, Adv. Theor. Math. Phys. 2 
(1998) 253, 
\hepth{9802150}; E. Witten, {\it Baryons and branes in anti de Sitter 
space}, JHEP {\bf 9807}, 006 (1998), \hepth{9805112}.

\bibitem{aharony}  O. Aharony, S. Gubser, J. Maldacena, 
H. Ooguri, Y. Oz, 
Phys. Rept. 323 (2000) 183;\\
G.T. Horowitz and J. Polchinski, gr-qc/0602037.

\bibitem{bl} J. Bagger and N. Lambert, \hepth{0611108}; \hepth{0711.0955}; 
\hepth{0712.3738}.

\bibitem{Gustav}
  A.~Gustavsson,
  arXiv:0709.1260 [hep-th]; A.~Gustavsson,
  JHEP {\bf 0804}, 083 (2008)
  [arXiv:0802.3456 [hep-th]].
\bibitem{Bandres:2008vf}
  M.~A.~Bandres, A.~E.~Lipstein and J.~H.~Schwarz,
  JHEP {\bf 0805}, 025 (2008)
  [arXiv:0803.3242 [hep-th]].



\bibitem{Mukhi:2008ux}
  S.~Mukhi and C.~Papageorgakis,
  JHEP {\bf 0805}, 085 (2008)
  [arXiv:0803.3218 [hep-th]].
\bibitem{Morozov:2008cb}
  A.~Morozov,
  JHEP {\bf 0805}, 076 (2008)
  [arXiv:0804.0913 [hep-th]].

\bibitem{Distler:2008mk}
  J.~Distler, S.~Mukhi, C.~Papageorgakis and M.~Van Raamsdonk,
  JHEP {\bf 0805}, 038 (2008)
  [arXiv:0804.1256 [hep-th]].
\bibitem{Gran:2008vi}
  U.~Gran, B.~E.~W.~Nilsson and C.~Petersson,
  arXiv:0804.1784 [hep-th].
\bibitem{Gaiotto:2008sd}
  D.~Gaiotto and E.~Witten,
N=4
  arXiv:0804.2907 [hep-th].

\bibitem{Ho:2008nn}
  P.~M.~Ho and Y.~Matsuo,
  JHEP {\bf 0806}, 105 (2008)
  [arXiv:0804.3629 [hep-th]].

\bibitem{verlinde}
  S.~Benvenuti, D.~Rodriguez-Gomez, E.~Tonni and H.~Verlinde,
  arXiv:0805.1087 [hep-th].

\bibitem{Ho:2008ei}
  P.~M.~Ho, Y.~Imamura and Y.~Matsuo,
  JHEP {\bf 0807}, 003 (2008)
  [arXiv:0805.1202 [hep-th]].

\bibitem{Honma:2008un}
  Y.~Honma, S.~Iso, Y.~Sumitomo and S.~Zhang,
  arXiv:0805.1895 [hep-th].
\bibitem{Song:2008bi}
  Y.~Song,
  arXiv:0805.3193 [hep-th].
\bibitem{shamik}
  S.~Banerjee and A.~Sen,
  arXiv:0805.3930 [hep-th].

\bibitem{Hosomichi:2008jd}
  K.~Hosomichi, K.~M.~Lee, S.~Lee, S.~Lee and J.~Park,
  arXiv:0805.3662 [hep-th].
\bibitem{Lin:2008qp}
  H.~Lin,
  arXiv:0805.4003 [hep-th].
\bibitem{Bandres:2008kj}
  M.~A.~Bandres, A.~E.~Lipstein and J.~H.~Schwarz,
  arXiv:0806.0054 [hep-th].

\bibitem{Park:2008qe}
  J.~H.~Park and C.~Sochichiu,
  arXiv:0806.0335 [hep-th].
\bibitem{abjm}
  O.~Aharony, O.~Bergman, D.~L.~Jafferis and J.~Maldacena,
  ``N=6 superconformal Chern-Simons-matter theories, M2-branes and their
  gravity duals,''
  arXiv:0806.1218 [hep-th].
\bibitem{Benna:2008zy}
  M.~Benna, I.~Klebanov, T.~Klose and M.~Smedback,
Correspondence,''
  arXiv:0806.1519 [hep-th].
\bibitem{Ezhuthachan:2008ch}
  B.~Ezhuthachan, S.~Mukhi and C.~Papageorgakis,
  JHEP {\bf 0807}, 041 (2008)
  [arXiv:0806.1639 [hep-th]].

\bibitem{Cecotti:2008qs}
  S.~Cecotti and A.~Sen,
  arXiv:0806.1990 [hep-th].

\bibitem{Mauri:2008ai}
  A.~Mauri and A.~C.~Petkou,
  arXiv:0806.2270 [hep-th].
\bibitem{Nishioka:2008gz}
  T.~Nishioka and T.~Takayanagi,
  arXiv:0806.3391 [hep-th].
\bibitem{Honma:2008jd}
  Y.~Honma, S.~Iso, Y.~Sumitomo and S.~Zhang,
  arXiv:0806.3498 [hep-th].

\bibitem{Bhattacharya:2008bja}
  J.~Bhattacharya and S.~Minwalla,
  arXiv:0806.3251 [hep-th].
\bibitem{Hosomichi:2008jb}
  K.~Hosomichi, K.~M.~Lee, S.~Lee, S.~Lee and J.~Park,
Orbifolds,''
  arXiv:0806.4977 [hep-th].
\bibitem{Bagger:2008se}
  J.~Bagger and N.~Lambert,
  arXiv:0807.0163 [hep-th].
\bibitem{Terashima:2008sy}
  S.~Terashima,
  arXiv:0807.0197 [hep-th].
\bibitem{Grignani:2008te}
  G.~Grignani, T.~Harmark, M.~Orselli and G.~W.~Semenoff,
  arXiv:0807.0205 [hep-th];
  G.~Grignani, T.~Harmark and M.~Orselli,
  arXiv:0806.4959 [hep-th].
\bibitem{Bandres:2008ry}
  M.~A.~Bandres, A.~E.~Lipstein and J.~H.~Schwarz,
  arXiv:0807.0880 [hep-th].

\bibitem{Zhou:2008sa}
  Y.~Zhou,
  arXiv:0807.0890 [hep-th].
\bibitem{Gomis:2008vc}
  J.~Gomis, D.~Rodriguez-Gomez, M.~Van Raamsdonk and H.~Verlinde,
  arXiv:0807.1074 [hep-th].
\bibitem{Kim:2008gn}
  N.~Kim,
  arXiv:0807.1349 [hep-th].
\bibitem{Pang:2008hw}
  Y.~Pang and T.~Wang,
  arXiv:0807.1444 [hep-th].
\bibitem{Garousi:2008ik}
  M.~R.~Garousi, A.~Ghodsi and M.~Khosravi,
  arXiv:0807.1478 [hep-th].

\bibitem{Hashimoto:2008iv}
  A.~Hashimoto and P.~Ouyang,
  arXiv:0807.1500 [hep-th].

\bibitem{Verlinde:2008di}
  H.~Verlinde,
  arXiv:0807.2121 [hep-th].

\bibitem{Lee:2008ui}
  B.~H.~Lee, K.~L.~Panigrahi and C.~Park,
  arXiv:0807.2559 [hep-th].
\bibitem{Krishnan:2008zm}
  C.~Krishnan and C.~Maccaferri,
  JHEP {\bf 0807}, 005 (2008)
  [arXiv:0805.3125 [hep-th]];
  C.~Krishnan,
  arXiv:0807.4561 [hep-th].

\bibitem{cveticcp3}
  M.~Cvetic, H.~Lu and C.~N.~Pope,
  Nucl.\ Phys.\  B {\bf 597}, 172 (2001)
  [arXiv:hep-th/0007109].
 
\bibitem{eh} T. Eguchi and A.J. Hanson, Ann. Phys. 120 (1979) 82;
T. Eguchi and A.J. Hanson, Phys. Lett. B74 (1978) 249.

\bibitem{zt} L.~A.~Pando Zayas and A.~A.~Tseytlin,
  ``3-branes on spaces with R x S(2) x S(3) topology,''
  Phys.\ Rev.\  D {\bf 63}, 086006 (2001)
  [arXiv:hep-th/0101043]. 
\bibitem{ks} I.~R.~Klebanov and M.~J.~Strassler,
  JHEP {\bf 0008}, 052 (2000)
  [arXiv:hep-th/0007191].
\bibitem{cvetic1} W. Chen, M. Cvetic, H. L\"u, C.N. Pope and J.F. 
Vazquez-Poritz, {Resolved Calabi-Yau Cones and Flows from $L^{abc}$ 
Superconformal Field Theories}, \hepth{0701082}.
\bibitem{hs} H. Singh, { 3-Branes on Eguchi-Hanson 6D Instantons},
Gen. Rel. Grav, {\bf 39} (2007) 839; 
\hepth{0701140}.
\bibitem{km} I.R. Klebanov and A. Murugan, {Gauge/Gravity 
Duality and Warped Resolved Conifold}, \hepth{0701064}.

\bibitem{cvetic}
  M.~Cvetic, G.~W.~Gibbons, H.~Lu and C.~N.~Pope,
  Commun.\ Math.\ Phys.\  {\bf 232}, 457 (2003)
  [arXiv:hep-th/0012011];
  M.~Cvetic, G.~W.~Gibbons, H.~Lu and C.~N.~Pope,
  Nucl.\ Phys.\  B {\bf 617}, 151 (2001)
  [arXiv:hep-th/0102185].


\end{thebibliography}
\end{document}